\documentclass[12pt]{article}
\usepackage{graphicx}
\usepackage{hyperref}
\usepackage{amsmath}
\usepackage{amscd,amssymb}
\usepackage{xcolor}

\newcommand{\cF}{\mathcal{F}}

\newcommand{\cG}{\mathcal{G}}
\newcommand{\fz}{\mathfrak{z}}

\def\cR{{\mathcal R}}

\def\cB{{\mathcal B}}

\def\cF{{\mathcal F}}

\def\cG{{\mathcal G}}

\def\cQ{{\mathcal Q}}

\def\mR{{\mathfrak R}}
\def\mT{{\mathfrak T}}

\def\mC{{\mathfrak C}}

\def\mS{{\mathfrak S}}

\newtheorem{theorem}{Theorem}[section]

\newtheorem{definition}{Definition}[section]

\newcommand{\IK}{\mathbb{K}}

\newcommand{\beq}{\begin{eqnarray}}
\newcommand{\eeq}{\end{eqnarray}}
\numberwithin{equation}{section}

\begin{document}

\begin{center}
{\large\bf  Elliptic Genera and $q$-Series Development
in Analysis, String Theory, and {\it N=2} Superconformal Field Theory}
\end{center}


\vspace{0.1in}

\begin{center}
{\large
L. Bonora $^{(a)}$
\footnote{E-mail: bonora@sissa.it},
A. A. Bytsenko $^{(b)}$
\footnote{E-mail: aabyts@gmail.com},
M. Chaichian $^{(c)}$
\footnote{E-mail: masud.chaichian@helsinki.fi}

and A. E. Gon\c{c}alves $^{(b)}$
\footnote{E-mail: aedsongoncalves@gmail.com}}

\vspace{5mm}
$^{(a)}$
{\it International School for Advanced Studies (SISSA/ISAS) \\
Via Bonomea 265, 34136 Trieste and INFN, Sezione di Trieste, Italy}

\vspace{0.2cm}
$^{(b)}$
{\it
Departamento de F\'{\i}sica, Universidade Estadual de
Londrina\\ Caixa Postal 6001,
Londrina-Paran\'a, Brazil}

\vspace{0.2cm}
$^{(c)}$
{\it
Department of Physics, University of Helsinki\\
P.O. Box 64, FI-00014 Helsinki, Finland}

\end{center}

\vspace{0.1in}

\begin{abstract}
In this article we examine the Ruelle type spectral functions $\cR(s)$,
which define an overall description of the content of the work.
We investigate the Gopakumar-Vafa reformulation of the string partition
functions, describe the ${\it N=2}$ Landau-Ginzburg
model in terms of Ruelle type spectral functions. Furthermore, we
discuss the basic properties satisfied
by elliptic genera in ${\it N=2}$ theories, construct the functional
equations for $\cR(s)$, and analyze the modular transformation laws
for the elliptic genus of the Landau-Ginzburg model and study
their properties in details.
\end{abstract}

\vspace{0.1in}

\begin{flushleft}
MSC \, 05A30 (q-Calculus and related topics)

\vspace{0.3in}
April 2019
\end{flushleft}

\newpage

\tableofcontents

\section{Introduction}

In this article we shall make considerable use of Ruelle type functions $\cR(s)$, which
should give a well-balanced description of the content of the whole article.
These functions are connected to symmetric functions (so-called S-functions $s_\lambda(x)$); the theory
of S-functions was developed by Schur \cite{Schur}. These functions play important role in the representation
theory of finite-dimensional classical Lie algebras.
Functions ${\mathcal R}(s)$ are an alternating product of more complicate factors, each of which are so-called
Patterson-Selberg zeta-functions $Z_{\Gamma}$ \cite{Williams}.

Now let us consider three-geometry with an orbifold description $H^3/\Gamma$. The complex unimodular group $G=SL(2, {\mathbb C})$
acts on the real hyperbolic three-space $H^3$ in a standard way, namely for $(x,y,z)\in H^3$ and $g\in G$, one gets
$g\cdot(x,y,z)= (u,v,w)\in H^3$. Thus for $r=x+iy$,\,
$g= \left[ \begin{array}{cc} a & b \\ c & d \end{array} \right]$,
$
u+iv = [(ar+b)\overline{(cr+d)}+ a\overline{c}z^2]\cdot
[|cr+d|^2 + |c|^2z^2]^{-1},\,
w = z\cdot[
{|cr+d|^2 + |c|^2z^2}]^{-1}\,.
$
Here the bar denotes the complex conjugation. Let $\Gamma \in G$ be the discrete group of $G$
defined as
\begin{eqnarray}
\Gamma & = & \{{\rm diag}(e^{2n\pi ({\rm Im}\,\tau + i{\rm Re}\,\tau)},\,\,  e^{-2n\pi ({\rm Im}\,\tau + i{\rm Re}\,\tau)}):
n\in {\mathbb Z}\}
= \{{\mathfrak g}^n:\, n\in {\mathbb Z}\}\,,
\nonumber \\
{\mathfrak g} & = &
{\rm diag}(e^{2\pi ({\rm Im}\,\tau + i{\rm Re}\,\tau)},\,\,  e^{-2\pi ({\rm Im}\,\tau + i{\rm Re}\,\tau)})\,.
\end{eqnarray}
One can define a Selberg-type zeta function for the group
$\Gamma = \{{\mathfrak g}^n : n \in {\mathbb Z}\}$ generated by a single hyperbolic element of the form
${\mathfrak g} = {\rm diag}(e^z, e^{-z})$, where $z=\alpha+i\beta$ for $\alpha,\beta >0$. In fact, we will take
$\alpha = 2\pi {\rm Im}\,\tau$, $\beta= 2\pi {\rm Re}\,\tau$.
For the standard action of $SL(2, {\mathbb C})$ on $H^3$ one has
\begin{equation}
{\mathfrak g}
\left[ \begin{array}{c} x \\ y\\ z \end{array} \right]
=
\left[\begin{array}{ccc} e^{\alpha} & 0 & 0\\ 0 & e^{\alpha} & 0\\ 0
& 0 & \,\,e^{\alpha} \end{array} \right]
\left[\begin{array}{ccc} \cos(\beta) & -\sin (\beta) & 0\\
\sin (\beta) & \,\,\,\,\cos (\beta) & 0
\\ 0 & 0 & 1 \end{array} \right]
\left[\begin{array}{c} x \\ y\\ z \end{array} \right]
\,.
\end{equation}
Therefore, ${\mathfrak g}$ is the composition of a rotation in ${\mathbb R}^2$ with complex eigenvalues
$\exp (\pm i\beta)$ and a dilatation $\exp (\alpha)$. There exists the Patterson-Selberg spectral function $Z_\Gamma (s)$,
meromorphic on $\mathbb C$, given for ${\rm Re}\, s> 0$ by
the formula \cite{Bytsenko}
\begin{equation}
{\rm log}\, Z_{\Gamma} (s)   =
-\frac{1}{4}\sum_{n=1}^{\infty}\frac{e^{-n\alpha(s-1)}}
{n[\sinh^2\left(\frac{\alpha n}{2}\right)
+\sin^2\left(\frac{\beta n}{2}\right)]}\,.
\label{logZ}
\end{equation}
The Patterson-Selberg function can be attached to ${H}^3/\Gamma$ as follows :
\begin{equation}
Z_\Gamma(s) :=\prod_{k_1, k_2 \in \mathbb{Z}_+ \cup \{0\}} [1-(e^{i\beta})^{k_1}(e^{-i\beta})^{k_2}e^{-(k_1+k_2+s)\alpha}]\,.
\label{zeta00}
\end{equation}
Zeros of $Z_\Gamma (s)$ are the complex numbers
\begin{equation}
\zeta_{n,k_{1},k_{2}} = -\left(k_{1}+k_{2}\right)+i\left(k_{1}-
k_{2}\right)\beta/\alpha+ 2\pi  in/\alpha\,\,\, (n \in {\mathbb Z}).
\end{equation}
In our applications we shall  consider a compact hyperbolic three-manifold
$G/\Gamma$ with $G = SL(2, {\mathbb C})$.
By combining the characteristic class representatives of field theory elliptic
genera one can compute quantum partition functions in terms of the
spectral functions of hyperbolic three-geometry, we shall demonstrate that in the
course of this article.

Let us introduce next Ruelle type spectral functions ${\mathcal R}(s)$ associated with
hyperbolic three-geometry \cite{BB}, which can be continued meromorphically
to the entire complex plane $\mathbb C$. Let $\chi$ be an orthogonal representation of $\pi_1(X)$.
Using the Hodge decomposition, the vector space $H(X;\chi)$ of twisted
cohomology classes can be embedded into $\Omega(X;\chi)$ as the
space of harmonic forms. This embedding induces a norm
$|\cdot|^{RS}$ on the determinant line ${\rm det}H(M;\chi)$. The
Ray-Singer norm $||\cdot||^{RS}$ on ${\rm det}H(X;\chi)$ is
defined by \cite{Ray}
\begin{equation}
||\cdot||^{RS}\stackrel{def}=|\cdot|^{RS}\prod_{p=0}^{{\rm dim}\,X}
\left[\exp\left(-\frac{d}{ds}
\zeta (s|L_p)|_{s=0}\right)\right]^{(-1)^pp/2}
\mbox{,}
\end{equation}
where the zeta function $\zeta (s|L_p)$ of the Laplacian
acting on the space of
$p$-forms orthogonal to the harmonic forms has been used. For a
closed connected orientable smooth manifold of odd dimension
and for Euler structure
$\eta\in {\rm Eul}(X)$, the Ray-Singer norm of its cohomological
torsion $\tau_{an}(X;\eta)=\tau_{an}(X)\in {\rm det}H(X;\chi)$ is
equal to the positive
square root of the absolute value of the monodromy of $\chi$
along the characteristic class $c(\eta)\in H^1(X)$ \cite{Farber}:
$||\tau_{an}(X)||^{RS}=|{\rm det}_{\chi}c(\eta)|^{1/2}$.
In the special
case where the flat bundle $\chi$ is acyclic, we have
\begin{equation}
\left[\tau_{an}(X)\right]^2
=|{\rm det}_{\chi}c(\eta)|
\prod_{p=0}^{{\rm dim}\,X}\left[\exp\left(-\frac{d}{ds}
\zeta (s|L_p)|_{s=0}\right)\right]^{(-1)^{p+1}p}
\mbox{.}
\label{RS}
\end{equation}
For a  closed oriented hyperbolic three-manifolds of the form
$X = {H}^3/\Gamma$, and for acyclic
$\chi$, the $L^2$-analytic torsion has the form
\cite{Fried,Bytsenko97}:
$[\tau_{an}(X)]^2={\mathcal R}(0)$, where
${\mathcal R}(s)$ is the Ruelle function
(it can be continued meromorphically to the entire complex plane
$\mathbb C$).
The function ${\mathcal R}(s)$ is an alternating
product of more complicate factors, each of which is a Selberg type zeta function $Z_{\Gamma}(s)$
(see Sect. \ref{Ruelle})

\subsection{Structure of the article and our key results}

We begin in Sect. \ref{algebra} with algebra of power series (of commuting variables). Then the polynomial
ring of symmetric functions $\Lambda(X)$ over $X$ we analyze in subsect. \ref{ring}. Algebraic properties
of $\Lambda(X)$, including plethysms and the Cauchy-Binet formulas, we examine in detail in subsect. \ref{lambda}.

At present q-series have reappeared in quantum physics and mathematics -- Lie
algebras, statistics, transcendental number theory, additive number theory and
in classical analysis. In this paper we apply q-series approach to several
interesting physical models, as in
Sections \ref{q-series} and Sect. \ref{N=2}.

We discuss the infinite specializations and $q$-series in Sect. \ref{q-series}.
Then the theory of Ruelle type functions $\cR(s)$, which are an alternating product of more complicate factors,
each of them is the Patterson-Selberg zeta function,
with its connection to Euler series is developed in subsect. \ref{Ruelle}.
Explicit formulas for the refined-shifted topological vertex are deduced in terms of Ruelle type spectral
functions in subsect. \ref{vertex}.
In subsection \ref{GV} we investigate the Gopakumar-Vafa reformulation for string partition functions in terms of
Ruelle type spectral functions.

Finally in Sect. \ref{N=2} we analyze the $\it N=2$ superconformal field theory. First of all we investigate
Bailey's transform and infinite hierarchy of Bailey's chain, including Ruelle type functions, subsection
\ref{Bailey}. Then in subsection \ref{RR} we begin by showing how some Rogers-Ramanujan identities can be
rewrite in term of $\cR(s)$.
The ${\it N=2}$ Landau-Ginzburg model we describe and investigate in subsect. \ref{LG}. The relation
between ${\it N=2}$ minimal models and Landau-Ginzburg theories has been proposed by E. Witten \cite{Witten}.
We discuss the basic properties satisfied by elliptic genera in ${\it N=2}$ theories. We construct the functional
equations for $\cR(s)$ functions and analyze the modular transformation laws for the elliptic genus of the
Landau-Ginzburg model.

\section{Algebra of power series (of commuting variables)}
\label{algebra}

Let $W$ be a set of all
sequences $\{s_n\}_{n=1}^\infty$ with integral non-negative terms. Let for each sequence
$s\in W$ the correspond number $\alpha_s$.

\begin{definition}
By definition such correspondence assigne a formal power series with coefficients
$\sum_{s\in W}\alpha_s \prod_{n=1}^\infty x_n^{s_n}$; $x_1, x_2, \ldots$.
Define
\begin{eqnarray}
\!\!\!\!\!\!\!\!\!\!\!\!
&&\left( \sum_{s\in W}\alpha_s\prod_{n=1}^\infty x_n^{s_n}\right)
+ \left( \sum_{s\in W}\beta_s\prod_{n=1}^\infty x_n^{s_n}\right)
\stackrel{def}{=} \sum_{s\in W}(\alpha_s+\beta_s)\prod_{n=1}^\infty x_n^{s_n};
\nonumber \\
\!\!\!\!\!\!\!\!\!\!\!\!
&&\beta\sum_{s\in W}\alpha_s\prod_{n=1}^\infty x_n^{s_n}\stackrel{def}{=}
\sum_{s\in W}\beta\alpha_s\prod_{n=1}^\infty x_n^{s_n},\,\,\,\,\ \beta\in \IK,
\nonumber \\
\!\!\!\!\!\!\!\!\!\!\!\!
&&\left( \sum_{s\in W}\alpha_s\prod_{n=1}^\infty x_n^{s_n}\right)
\left( \sum_{s\in W}\beta_s\prod_{n=1}^\infty x_n^{s_n}\right)
\stackrel{def}{=}\sum_{s\in W}\gamma_s\prod_{n=1}^\infty x_n^{s_n}.
\end{eqnarray}
Where
$
 \gamma_s = \sum_{s^{\prime}, s^{\prime \prime}}\alpha_{s^\prime} \beta_{s^{\prime\prime}},
$
the summation is realized for pair of sequances which satisfy the condition
$s_n^\prime + s_n^{\prime\prime} = s_n$ for any natural $n$.
\end{definition}

\subsection{Polynomial ring of symmetric functions}
\label{ring}

Our aim in this section is to exploit the Hopf algebra of the ring $\Lambda(X)$
of symmetric functions of the independent variables $(x_1, x_2, \ldots)$,
finite or contably infinite in number, that constitute the alphabet $X$.

Let ${\mathbb Z}[x_1,\ldots,x_n]$ be the polynomial ring, or the
ring of formal power series, in $n$ commuting variables $x_1,\ldots,x_n$. The symmetric group
$S_n$ acts on this ring by permuting the variables.
For $\pi\in S_n$ and $f \in {\mathbb Z}[x_1,\ldots,x_n]$ we have
$
\pi f(x_1,\ldots,x_n) = f(x_{\pi(1)},\ldots,x_{\pi(n)}).
$
We are interested in the subring of functions invariant under this
action, $\pi f = f$, that is to say the ring of symmetric polynomials in
$n$ variables:
$
\Lambda(x_1,\ldots,x_n) = {\mathbb Z}[x_1,\ldots,x_n]^{S_n}.
$
This ring may be graded by the degree of the polynomials, so that
$
\Lambda(X)=\oplus_n\ \Lambda^{(n)}(X)
$,
where $\Lambda^{(n)}(X)$ consists of homogenous symmetric polynomials  in $x_1,\ldots,x_n$
of total degree $n$.

In order to work with an arbitrary number of variables, following
Macdonald~\cite{Macdonald}, we define the ring of symmetric functions
$\Lambda = \lim_{n\rightarrow\infty}\Lambda(x_1,\ldots,x_n)$ in its stable limit  ($n\rightarrow
\infty$). There exist various bases of $\Lambda(X)$:

A $\mathbb Z$ basis of $\Lambda^{(n)}$ is provided by the
monomial symmetric functions $\{m_\lambda\}$, where $\lambda$ is any partitions of $n$.

The other (integral and rational) bases for $\Lambda^{(n)}$ are indexed by the
partitions $\lambda$ of $n$. There are the complete,
elementary and power sum symmetric functions bases defined multiplicatively in terms of
corresponding one part functions by:
$h_{\lambda}=h_{\lambda_1}h_{\lambda_2}\cdots h_{\lambda_n}$,
$e_{\lambda}=e_{\lambda_1}e_{\lambda_2}\cdots e_{\lambda_n}$ and
$p_{\lambda}=p_{\lambda_1}p_{\lambda_2}\cdots p_{\lambda_n}$ where the one part functions
are defined for $\forall n \in {\mathbb Z}_+$ by
\begin{equation}
h_n(X) = \sum_{i_1\leq i_2\cdots\leq i_n}x_{i_1}x_{i_2}\cdots x_{i_n},\,\,\,\,\,\,\,\,
e_n(X) = \sum_{i_1<i_2\cdots<i_n}x_{i_1}x_{i_2}\cdots x_{i_n},\,\,\,\,\,\,\,\,
p_n(X) = \sum_{i}x_i^n,
\end{equation}

with the convention $h_0 = e_0 = p_0 =1,\, h_{-n} = e_{-n} = p_{-n} = 0$.
Three of these bases are multiplicative,
with $h_{\lambda}=h_{\lambda_1}h_{\lambda_2}\cdots h_{\lambda_n}$,
$e_{\lambda}=e_{\lambda_1}e_{\lambda_2}\cdots e_{\lambda_n}$ and
$p_{\lambda}=p_{\lambda_1}p_{\lambda_2}\cdots p_{\lambda_n}$.
The relationships between the various bases we just mention at this stage by
the transitions
\begin{equation}
p_\rho(X) = \sum_{\lambda\,\vdash n}\chi_\rho^\lambda s_\lambda(X)
\,\,\,\,\,\,\, {\rm and} \,\,\,\,\,\,\,
s_\lambda(X) = \sum_{\rho\,\vdash n}\ \fz_\rho^{-1}\ \chi^\lambda_\rho\ p_\rho(X)\,.
\label{Eq-p-s}
\end{equation}
For each partition $\lambda$, the Schur function is defined by
\begin{equation}
s_\lambda(X)\equiv s_\lambda(x_1,x_2, \ldots, x_n) = \frac{\sum_{\sigma\in S_n}{\rm sgn}
(\sigma)X^{\sigma(\lambda+\delta)}}{\prod_{i<j}(x_i-x_j)}\,,
\end{equation}
where $\delta = (n-1,n-2,\ldots,1,0)$. In fact both $h_n$ and $e_n$ are special Schur functions,
$h_n = s_{(n)},\, e_n = s_{(1^n)}$, and their generating functions are expressed in terms of
the power-sum $p_n$:
\begin{equation}
\sum_{n\geq 0}h_nz^n = \exp (\sum_{n=1}^\infty(p_n/n)z^n), \,\,\,\,\,\,\,\,\,\,
\sum_{n\geq 0}e_nz^n = \exp (-\sum_{n=1}^\infty(p_n/n)(-z)^n)\,.
\end{equation}
The Jacobi-Trudi formula \cite{Macdonald} express the Schur functions in terms of $h_n$ or
$e_n$: $s_\lambda = {\rm det}(h_{\lambda_i-i+j}) = {\rm det}(e_{\lambda^\prime-i+j})$,
where $\lambda^\prime$ is the conjugate of $\lambda$.
An involution $\iota: \Lambda\rightarrow \Lambda$ can be defined by $\iota(p_n) = (-1)^{n-1}p_n$.
Then it follows that
$\iota(h_n) = e_n$. Also we have $\iota(s_\lambda) = s_{\lambda^\prime}$.
$\chi^\lambda_\rho$ is the character of the irreducible representation
of the symmetric groups $S_n$ specified by $\lambda$ in the conjugacy class
specified by $\rho$. These characters satisfy the orthogonality conditions
\begin{eqnarray}
\sum_{\rho\,\vdash n}\ \fz_\rho^{-1}\ \chi^\lambda_\rho\ \chi^\mu_\rho = \delta_{\lambda,\mu}
\,\,\,\,\,\,\, {\rm and} \,\,\,\,\,\,\,
\sum_{\lambda\,\vdash n}\ \fz_\rho^{-1} \chi^\lambda_\rho \chi^\lambda_\sigma\ =
\delta_{\rho,\sigma}\,.
\label{Eq-chi-orth}
\end{eqnarray}
The significance of the Schur function basis lies in the fact that with respect
to the usual Schur-Hall scalar product
$\langle \cdot \,|\, \cdot \rangle_{\Lambda(X)}$ on $\Lambda(X)$ we have
\begin{eqnarray}
\langle s_\mu(X) \,|\, s_\nu(X) \rangle_{\Lambda(X)}
= \delta_{\mu,\nu}\,\,\,\,\,\, {\rm and}\,\,\,\,\, {\rm therefore}
\,\,\,\,\,
\langle p_\rho(X) \,|\, p_\sigma(X) \rangle_{\Lambda(X)}
= \fz_\rho \delta_{\rho,\sigma}\,,
\label{Eq-scalar-prod-p}
\end{eqnarray}
where $\fz_\lambda = \prod_i i^{m_i}m_i!$ for $\lambda = (1^{m_1}, 2^{m_2}, \cdots)$.

\subsection{Algebraic properties of $\Lambda(X)$}
\label{lambda}

The ring $\Lambda(X)$ of symmetric functions over $X$ has a Hopf algebra
structure, and there are two further algebraic and two coalgebraic operations.
For notation and basic properties we refer the reader to
\cite{fauser:jarvis:2003a,fauser:jarvis:king:wybourne:2006a,Fauser10} and
references therein.

Plethysms (compositions) are denoted by $\circ$ or by
means of square brackets $[\,\,]$; plethysm coproduct is denoted by $\triangledown$.
The corresponding coproduct maps are specified by $\Delta$ for the outer coproduct.
Notation $\delta$ we use for the inner coproduct.

The coproduct coefficients themselves are obtained
from the products by duality using the Schur-Hall scalar product and the
self-duality of $\Lambda(X)$. For example, for all $A,B\in\Lambda(X)$:
\begin{eqnarray}
A\circ B  =  A[B];\,\,\,\,\,  \Delta(A) &=& A_{(1)}\otimes A_{(2)};
\nonumber \\
\delta(A) = A_{[1]}\otimes A_{[2]};\,\,\,\,\,\triangledown (A) &=& A_{\langle 1\rangle}\otimes A_{\langle 2\rangle}\,.
\end{eqnarray}

In terms of the Schur function basis $\{s_\lambda\}_{\lambda\,\vdash
n,n\in\mathbb{Z}_+}$ the product and coproduct maps give
rise to the particular sets of coefficients specified as follows:
\begin{eqnarray}
s_\mu s_\nu & = & \sum_\lambda c^\lambda_{\mu,\nu}s_\lambda \,;
\,\,\,\,\,\,\,\,\,
\Delta(s_\lambda)
= s_{\lambda_{( 1 ) }}\otimes s_{\lambda_{( 2 )}}
=\sum_{\mu,\nu}c^\lambda_{\mu,\nu} s_\mu\otimes s_\nu \,;
\nonumber \\
s_\mu\ast s_\nu & = & \sum_\lambda g^\lambda_{\mu,\nu}s_\lambda \,;
\,\,\,\,\,\,\,\,\,
\delta(s_\lambda) = s_{\lambda_{[1]}}\otimes s_{\lambda_{[2]}}\,\,\,\,\,
=\sum_{\mu,\nu} g^\lambda_{\mu,\nu}s_\mu\otimes s_\nu\,;
\nonumber \\
s_\mu[s_\nu] & = & \sum_\lambda p_{\mu, \nu}^\lambda s_\lambda \,;
\,\,\,\,\,\,\,\,\,
\triangledown (s_\lambda) \,\, =  s_{\lambda_{\langle 1\rangle}}\otimes
s_{\lambda_{\langle 2\rangle}} \, = \sum_{\mu,\nu}  p_{\mu, \nu}^\lambda s_\mu\otimes s_\nu\,.
\end{eqnarray}
Here the $c^\lambda_{\mu,\nu}$ are Littlewood-Richardson coefficients,
the $g^\lambda_{\mu,\nu}$ are Kronecker coefficients and the
$p^\lambda_{\mu,\nu}$ are plethysm coefficients. All these coefficients
are non-negative integers. The Littlewood-Richardson coefficients
can be obtained, for example, by means of the Littlewood-Richardson
rule~\cite{littlewood:richardson:1934a,littlewood:1950a} or the hive
model~\cite{buch:2000a}. The Kronecker coefficients may determined directly
from characters of the symmetric group or by exploiting the Jacobi-Trudi
identity and the Littlewood-Richardson rule, while
plethysm coefficients have been the subject of a variety methods of
calculation~\cite{littlewood:1950b,chen:garsia:remmel:1984a}.
Note that the above sums are finite, since
$c^\lambda_{\mu,\nu} \geq 0$ iff $\vert\lambda\vert=\vert\mu\vert +\vert\nu\vert$;
$g^\lambda_{\mu,\nu} \geq 0$ iff $\vert\lambda\vert=\vert\mu\vert =\vert\nu\vert$;
$p^\lambda_{\mu,\nu} \geq 0$ iff $\vert\lambda\vert=\vert\mu\vert\,\vert\nu\vert$.

The Schur-Hall scalar product may be used to define skew Schur functions
$s_{\lambda/\mu}$ through the identities
$
c^\lambda_{\mu,\nu}
= \langle s_\mu\, s_\nu \vert s_\lambda \rangle  = \langle s_\nu\,\vert s_\mu^\perp (s_\lambda)
\rangle = \langle s_\nu \vert s_{\lambda/\mu} \rangle\,,
$
so that
$
s_{\lambda/\mu} =  \sum_\nu\ c^\lambda_{\mu,\nu}\ s_\nu\,.
$

In what follows we shall make considerable use of several infinite series
of Schur functions. The most important of these are the mutually inverse
pair defined by
\begin{eqnarray}
 \cF(t;X) & = & \prod_{i\geq 1} (1-t\,x_i)^{-1} = \sum_{m\geq 0} h_m(X)t^m\,
\label{Eq-M},\\
\cG(t;X) & = & \prod_{i\geq 1} (1-t\,x_i)\,\,\,\,\,\,
= \, \sum_{m\geq 0}(-1)^m e_m(X) t^m\,,
\label{Eq-L}
\end{eqnarray}
where as Schur functions $h_m(X)=s_{(m)}(X)$ and $e_m(X)=s_{(1^m)}(X)$.
For convenience, in the case $t=1$ we write $\cF(1;X)=\cF(X)$ and $\cG(1;X)=\cG(X)$.

{\bf Note on plethysms.}
\label{Plethysm} Plethysms are defined as compositions whereby for any $A,B\in\Lambda(X)$;
the plethysm $A[B]$ is $A$ evaluated over an alphabet $Y$ whose letters are the
monomials of $B(X)$, with each letter repeated as many times as the
multiplicity of the corresponding monomial. For example, the Schur function plethysm
is defined by $s_\lambda[s_\mu](X) = s_\lambda(Y)$, where $Y=s_\mu(X)$.

For all $A,B,C\in \Lambda(X)$ we have the following rules, due to
Littlewood~\cite{littlewood:1950a}, for manipulating plethysms:

\begin{eqnarray}
\label{littlewood:plethysm}
(A+B)[C] & = & A[C] + B[C]\,; \,\,\,\,\,\,\,\,\,\, A[B + C] = A_{(1)}[B] A_{(2)}[C]\,;
\nonumber \\
(AB)[C]  & = & A[C]B[C]\,; \,\,\,\,\,\,\,\,\,\,\,\,\,\,\,\,\,\,\,\,\,\,\,\,\,
A[BC] = A_{[1]}[B]A_{[2]}[C]\,;
\nonumber \\
 A[B[C]]  & = & (A[B])[C]\,.
\end{eqnarray}
These rules enable us to evaluate plethysms not only of outer and inner
products but also of outer and inner coproducts.

{\bf The Cauchy-Binet formulas.}
It is often convenient to represent an alphabet in an additive manner
$X$, as itself an element of the ring $\Lambda(X)$ in the
sense that
$x_1+x_2+\cdots\ = h_1(X) = e_1(X) = p_1(X) = s_{(1)}(X).
$
As elements of $\Lambda(X)\otimes\Lambda(Y)$ we have
$X + Y = \sum_{j=1}x_j+\sum_{j=1}y_j$,\, $XY = \sum_jx_j\sum_jy_j$.
With this notation, the outer coproduct gives
\begin{eqnarray}
\Delta(\cF) & = & \cF_{(1)}\otimes \cF_{(2)} = \cF\otimes \cF;
\,\,\,\,\,\,\,\,\,\,
\cF(X\!\!+\!\!Y) = \prod_i (1-x_i)^{-1}\,\prod_j\, (1-y_j)^{-1}\,;
\nonumber \\
\Delta(\cG) & = & \cG_{(1)}\otimes \cG_{(2)} = \cG\otimes \cG;
\,\,\,\,\,\,\,\,\,\,\,\,\,\,\,\,\,\,\,\,\,
\cG(X\!\!+\!\!Y) = \prod_i (1-x_i)\, \prod_j\,(1-y_j) \,,
\nonumber
\end{eqnarray}
so that $\cF(X\!\!+\!\!Y)=\cF(X)\,\cF(Y)$ and $\cG(X\!\!+\!\!Y)=\cG(X)\,\cG(Y)$.
For the inner coproduct:
\begin{eqnarray}
  \delta(\cF)
     &=\cF_{[1]}\otimes \cF_{[2]};
\,\,\,\,\,\,\,\,\,\,
       \cF(XY)
     &= \prod_{i,j} (1-x_iy_j)^{-1}\,;
   \nonumber \\
  \delta(\cG)
     &=\cG_{[1]}\otimes \cG_{[2]};
\,\,\,\,\,\,\,\,\,\,
       \cG(XY)
     &= \prod_{i,j} (1-x_iy_j) \,.
     \nonumber
\end{eqnarray}
The expansions of the products on the right hand sides of these expressions is
effected remarkably easily by evaluating the inner coproducts on the left:
\begin{eqnarray}
\delta(\cF) & = & \sum_{k\geq 0}\, \delta(h_k)
= \sum_{k\geq 0} \sum_{\lambda\vdash k}\ s_\lambda \otimes s_\lambda\,;
  \nonumber \\
\delta(\cG) & = & \sum_{k\geq0}\,(-1)^k\,\delta(e_k)
= \sum_{k\geq0}\, (-1)^k\, \sum_{\lambda\vdash k}\
s_\lambda\otimes s_{\lambda^\prime}\,.
  \nonumber
\end{eqnarray}
This gives immediately the well known Cauchy and Cauchy-Binet formulas:
\begin{eqnarray}
\cF(XY) & = & \prod_{i,j}\ (1-x_i\,y_j)^{-1}
= \sum_\lambda\ s_\lambda(X)\, s_\lambda(Y)\,;
\label{Eq-Cauchy}
    \\
\cG(XY) & = & \prod_{i,j}\ (1-x_i\,y_j)
= \sum_\lambda\ (-1)^{|\lambda|} s_\lambda(X)\,s_{\lambda^\prime}(Y) \,.
\label{Eq-dual-Cauchy}
\end{eqnarray}
Generally speaking, for any $F(X)\in \Lambda(X)$ with dual $F^\perp(X)$, by
linearly extending the above result we have $F(X)\ \cF(XY) = F^\perp(Y)\,(\cF(XY))$.

\section{Infinite specializations and $q$-series}
\label{q-series}

The study of $q$-series has been extensively enriched due to success in investigation of $q$-analogs
of the classical special functions. The large amount of activity of $q$-series has been appeared
in Lie algebras, statistical and quantum mechanics, trancendental number theory, and computer algebra.

By considering infinite specializations, i.e. setting $X= (x_1, x_2, \ldots,x_r,\ldots)
= (q, q^2, \ldots q^r,\ldots)$, $q=e^{2\pi i\tau}$,
we present the following useful identities:
\begin{eqnarray}
\cF(q; XY) & = &  \prod_{i,j}(1-qx_iy_j)^{-1}  =  \sum_\alpha q^\alpha s_\alpha(X)s_\alpha(Y),
\nonumber \\
\cG(q; XY) & = & \prod_{i,j}(1-qx_iy_j) \,\,\,\, =
\,\,\, \sum_\alpha (-q)^{|\alpha|} s_\alpha(X)s_{\alpha^\prime}(Y).
\end{eqnarray}

\subsection{Spectral functions of hyperbolic three-geometry}
\label{Ruelle}

The most important Euler series can be represent in the form of the
Ruelle type (Patterson-Selberg) spectral function $\cR(s)$ of hyperbolic three-geometry, in fact:
\begin{eqnarray}
\prod_{n=\ell}^{\infty}(1- q^{an+\varepsilon})
& = & \prod_{p=0, 1}Z_{\Gamma}(\underbrace{(a\ell+\varepsilon)(1-i\varrho(\tau))
+ 1 -a}_s + a(1 + i\varrho(\tau)p)^{(-1)^p}
\nonumber \\
& = &
\cR(s = (a\ell + \varepsilon)(1-i\varrho(\tau)) + 1-a),
\\
\prod_{n=\ell}^{\infty}(1+ q^{an+\varepsilon})
& = &
\prod_{p=0, 1}Z_{\Gamma}(\underbrace{(a\ell+\varepsilon)(1-i\varrho(\tau)) + 1-a +
i\sigma(\tau)}_s
+ a(1+ i\varrho(\tau)p)^{(-1)^p}
\nonumber \\
& = &
\cR(s = (a\ell + \varepsilon)(1-i\varrho(\tau)) + 1-a + i\sigma(\tau))\,.
\end{eqnarray}
Here $q= \exp(2\pi i\tau)$, $\varrho(\tau) =
{\rm Re}\,\tau/{\rm Im}\,\tau$,
$\sigma(\tau) = (2\,{\rm Im}\,\tau)^{-1}$,
$a$ is a real number, $\varepsilon, b\in {\mathbb C}$, $\ell \in {\mathbb Z}_+$.

{\bf Various expansions different from power series expansion.}
Let us consider
\begin{equation}
\cQ(q) = \prod_{n=1}^\infty (1-q^n)^{-C_n} = 1+ \sum_{n=1}^\infty \cB_nq^n.
\label{expansion}
\end{equation}
Usualy $\cQ(q)$ associated wth some generating function, and $\cB_n$ is related
sequence that will be study.
\begin{theorem} (\cite{Andrews}, Theorem 10.3)
Let in Eq. (\ref{expansion}) $C_n$ and $\cB_n$are integers, then
\begin{eqnarray}
n\cB_n &=& \sum_{k=1}^nD_k\cB_{n-k},
\label{B}
\\
D_k &=& \sum_{d\vert k}dC_d.
\label{D}
\end{eqnarray}
 Note that if either sequence $C_n$ or $\cB_n$ is given, then the other is uniquely
 determined by (\ref{B}) and (\ref{D}).
\end{theorem}

For $C_n = c\cdot n, c= const.$ the following relation holds:
\begin{eqnarray}
\prod_{n=\ell}^\infty(1-q^{an+\varepsilon})^{cn} &= &\cR(s=(a\ell+\varepsilon)(1-i\varrho(\tau))+1-a)^{c\ell}
\nonumber \\
&\times & \!\!\prod_{n=\ell+1}^\infty \cR(s= (an+\varepsilon)(1-i\varrho(\tau))+ 1-a)^c.
\end{eqnarray}

\subsection{The refined-shifted topological vertex}
\label{vertex}

{\bf The standard topological vertex $\mC_{\lambda\mu\nu}(q)$.} For the particular case
(there is no boundary condition), i.e. $\lambda=\mu=\nu = \emptyset$, the vertex
$\mC_{\emptyset\emptyset\emptyset}(q)= \mC_3$ coincides with 3d-- MacMahon function
$\mC_3(q)= \prod_{n=1}^\infty(1-q^n)^{-n}$ \cite{Drissi}.
 \begin{equation}
 \prod_{n=1}^\infty(1-q^n)^{-n} = \cR(s=1-i\varrho(\tau))^{-1}
\prod_{n= 2}^{\infty}\cR(s=n(1-i\varrho(\tau)))^{-1}
\end{equation}

{\bf The refined topological vertex $\mR_{\lambda\mu\nu}(q,t)$.}
$\mR_{\lambda\mu\nu}(q,t)$ can be constructed as a two
parameter extension of $\mC_{\lambda\mu\nu}(q)$.
In the particular case $\lambda=\mu=\nu =\emptyset$, the vertex $\mR_{\lambda\mu\nu}(q,t)
= \mR_{\emptyset\emptyset\emptyset}(q,t)$ coincide with the refined 3d-MacMahon function.
\begin{equation}
 \mR_{\emptyset\emptyset\emptyset}(q,t) = \prod_{n,k =1}^\infty(1-q^{n-1}t^k)
 =\prod_{k=1}^\infty \cR(s= \varepsilon(1-i\varrho(\tau))),
\label{3d}
 \end{equation}
 where $\varepsilon = k{\rm log} t/2\pi i\tau$.
Finally in the case $t=q$ in Eq. (\ref{3d}) we obtain the standard $\mC_3(q)$ relation.

{\bf Shifted topological vertex $\mS_{\lambda\mu\nu}(q)$.} With generic boundary
conditions the shifted topological vertex $\mS_{\lambda\mu\nu}(q)$ is the shifted 3d
MacMahon $\mS_{\emptyset\emptyset\emptyset}(q) = \mS_3(q)$, the generating functional
of strict plane partitions $\mS_3(q) = \prod_{n=1}^\infty \left(\frac{1+q^n}{1-q^n}\right)^n$ \cite{Drissi}.

\begin{equation}
\mS_3(q) =  \frac{\cR(s=1-i\varrho(\tau)+i\sigma(\tau))}{\cR(s=1-i\varrho(\tau))}
\cdot \frac{\prod_{n=2}^\infty\cR(s=n(1-i\varrho(\tau)+i\sigma(\tau)))}
{\prod_{n=2}^\infty \cR(s=n(1-i\varrho(\tau)))}.
\end{equation}

{\bf Refining the shifted vertex $\mT_{\lambda\mu\nu}(q,t)$.} The refining version
$\mT_{\lambda\mu\nu}(q,t)$ of the shifted topological vertex $\mS_{\lambda\mu\nu}(q)$
is a two parameters $q$ and $t$ with boundary conditions given by strict 2d partitions
$\lambda, \mu$ and $\nu$ \cite{Drissi}. In addition, $\mT_{\lambda\mu\nu}(q,t)$
is non cyclic with respect to the permutations of the strict 2d partitions $\lambda,\mu,\nu$,
i.e. $\mT_{\lambda\mu\nu}\neq $ $\mT_{\mu\nu\lambda}$ $\neq \mT_{\nu\lambda\mu}$, and what
is more
\begin{equation}
 \mT_{\lambda\mu\nu}(q, q) = \mS_{\lambda\mu\nu}(q), \,\,\,\,\,
 \mT_{\emptyset\emptyset\emptyset}(q,t) = \mT_3(q,t).
\end{equation}
 \begin{equation}
 \mT_3 (q,t) =\prod_{j=1}^\infty\prod_{k=1}^\infty \left(\frac{1+q^{j-1}t^k}{1-q^{j-1}t^k}\right) =
\prod_{k=1}^\infty\frac{\cR(s= \varepsilon(1-i\varrho(\tau))+i\sigma(\tau))}
{\cR(s=\varepsilon(1-i\varrho(\tau)))},
 \end{equation}
recall that $\varepsilon = k{\rm log}t/2\pi i\tau$. In this section we have analyzed the refining and the shifting
properties of the standard topological vertex $\mC_{\lambda\mu\nu}(q)$, the refined topological vertex
$\mR_{\lambda\mu\nu}(q,t)$, the shifted topological vertex $\mS_{\lambda\mu\nu}(q)$, and the refining the shifted
vertex $\mT_{\lambda\mu\nu}(q,t)$, in terms of spectral functions $\cR(s)$.

\subsection{Gopakumar-Vafa reformulation of string partition functions}
\label{GV}

In order to prove the Gopakumar-Vafa conjecture \cite{GopakumarI,GopakumarII,Gopakumar99}, we need to rewrite the sums
over partition by means of infinite products. In this section we propose a spectral function reformulation for such an
infinite products.

Denote by $F$ the generating series of Gromov-Witten invariants of a Calabi-Yau three-fold $X$.
Intuitively we can counts the number of stable maps with {\em connected} domain curves to $X$ in any given
nonzero homology classes. However, because of the existence of automorphisms, one has to perform the weighted
count by dividing by the order of the automorphism groups (hence Gromov-Witten invariants are in general rational numbers).
Based on $M$-theory considerations, Gopakumar and Vafa \cite{GopakumarII} made a remarkable conjecture on the
structure of $F$, in particular, on its integral properties. More precisely, integers $n^g_{\Sigma}$ are conjectured
to exists such that
\begin{equation}
F =  \sum_{\Sigma \in H_2(X)-\{0\}} \sum_{g \geq 0} \sum_{k\in {\mathbb Z}_+} {k}^{-1}
n^g_{\Sigma} (2\sin (k\lambda/2))^{2g-2}Q^{k\Sigma}\,.
\label{Gopakumar}
\end{equation}
For given $\Sigma$, there are only finitely many nonzero $n^g_{\Sigma}$,
$Q^{\Sigma} = \exp(-\int_{\Sigma}\omega)$, the holomorphic curve $\Sigma\in H_2(X, {\mathbb Z}):= H_2(X)$ is given by
$\int_{\Sigma}\omega$, where $\omega$ is the K\"{a}hlerian form on $X$.
Let us regard $q = \exp (i\lambda)$, for some real $\lambda=2\pi \tau$, as an element of $SU(2)$ represented by
the diagonal matrix ${\rm diag}\,(q , q^{-1})$. The generating series of {\it disconnected} Gromov-Witten invariants
is given by the string partition function:
$Z = \exp F$.

{\bf Topological string amplitudes.} We shall present now the interpretation
of topological string amplitudes in the form of the generating functions of
the BPS degeneracies of wrapped M2-branes, which give rise to
particles in the five dimensional theory \cite{Hollowood}. First consider IIA strings compactified on
a CY3-fold X. In this case the theory on the transverse four dimensions has {\it N=2} supersynnetry.
In four dimensions this theory has F-terms which can be calculated exactly \cite{Antoniadis,Bershadsky}.

Let $F_g$ be the topological string amplitudes (Gopakumar). In the A-twisted topologicasl theory $F_g$ arise as
integrals over the genus $g$ moduli space of Riemann surfaces and related to the generating functions
of the genus $g$ Gromov-Witten invariants. The topological string amplitudes can be compactly organized
into the generating function  $F(\lambda_s) = \sum_{g=0}^\infty \lambda_s^{2g-2}F_g$, where $\lambda_s$
is the constant self-dual graviphoton strength. Interesting physical interpretation of the generating
function $F(\lambda_s)$ the reader can find in \cite{GopakumarI,GopakumarII}

For second quantized strings and A-model result (the Gopakumar-Vafa conjecture) can be reformulated as follows
(see \cite{Hollowood}):
\begin{equation}
Z = \prod_{\Sigma \in H_2(X)}\prod_j \prod_{k=-j}^j \,\,\prod_{m \in {\mathbb Z}_+ \cup\, \{0\}}
(1- q^{2k+m+1}Q^{\Sigma})^{(-1)^{2j+1}(m+1)N_{\Sigma}^g}.
\label{Z2}
\end{equation}
In Eq. (\ref{Z2}) $j = g/2$,\, $k=-j, -j+1, \dots, j-1, j$, $Q^\Sigma = e^{-T_\Sigma}$ and $q= e^{-i\lambda_s}$.
Expressions (\ref{Gopakumar}) and (\ref{Z2}) look
very much like {\it counting} the states in a Hilbert space for the case of $j_L=0$ BPS states \cite{GopakumarI}.
The partition function counts M2-branes \cite{Hollowood}. In addition the integrality of $Z$ has to be directly related
to the same integrality in $F$. One can interpret $Z$ as the partition function of a second quantized theory which is built
purely out of fields creating M2-branes.

In terms of the Ruelle type spectral functions the string partition function can be reformulated as
follows
\begin{eqnarray}
\!\!\!Z &= &\!\!\!\!\!\! \prod_{\Sigma \in H_2(X)}\prod_j \prod_{k=-j}^j \,\,\prod_{m \in {\mathbb Z}_+ \cup\, \{0\}}
\!\!\!(1- q^{2k+m+1}Q^{\Sigma})^{(-1)^{2j+1}N_{\Sigma}^gm}\cdot(1- q^{2k+m+1}Q^{\Sigma})^{(-1)^{2j+1}N_{\Sigma}^g}
\nonumber \\
&=& \!\!\!\!\! \prod_{\Sigma \in H_2(X)}\prod_j \prod_{k=-j}^j\cR(s= (2k+\varphi+1)(1-i\varrho(\tau)))
\nonumber \\
&\times& \!\!\prod_{n=\ell+1}^\infty\cR(s= (n+2k+\varphi+1)(1-i\varrho(\tau)))^{(2j+1)N_{\Sigma}^g},
\end{eqnarray}
where $\varphi = {\rm log}Q^{\Sigma}/2\pi i\tau= -\int_{\Sigma}\omega/2\pi i\tau$.

\section{The {\it N=2} superconformal field theory}
\label{N=2}

\subsection{Bailey's transform and infinite hierarchy of Bailey's chain}
\label{Bailey}

Let us begin with notations: $(a; q)_\infty := \prod_{m=0}^\infty (1-aq^m)$, and $(a; q)_n := (a; q)_\infty/(aq^n; q)_\infty$.
W.N. Bailey made the following observation \cite{Bailey} (see also \cite{Andrews}), which is known
as {\it Bailey's transform}.
\begin{theorem} (lectures \cite{Andrews}, Theorem 3.1)\label{Th}
Let for a suitable convergence conditions,
\begin{eqnarray}
\beta_n & = & \sum_{r=0}^n \alpha_ru_{n-r}v_{n+r},
\\
\gamma_n & =& \sum_{r=n}^\infty \delta_r u_{r-n}v_{r+n},
\end{eqnarray}
then
\begin{equation}
 \sum_{n=0}^\infty \alpha_n \gamma_n = \sum_{n=0}^\infty \beta_n \delta_n.
\end{equation}
\end{theorem}
{\it Proof}.
\begin{equation}
\sum_{n=0}^\infty\alpha_n\gamma_n = \sum_{n=0}^\infty\sum_{r=n}^\infty\alpha_n\delta_r u_{r-n}v_{r+n}
= \sum_{r=0}^\infty\sum_{n=0}^r\alpha_n\delta_r u_{r-n}v_{r+n} = \sum_{r=0}^\infty\delta_r\beta_r.
\end{equation}
Important application of the Bailey's transform and Bailey's Lemma the reader can found in \cite{Bailey}, $\S$\,4

Let for $n\geq 0$
\begin{equation}
 \beta_n = \sum_{r=0}^n\frac{\alpha_r}{(q;q)_{n-r}(aq;q)_{n+r}},
 \label{beta1}
 \end{equation}
then also
\begin{equation}
 {\beta^\prime}_n = \sum_{r=0}^n\frac{{\alpha^\prime}_r}{(q;q)_{n-r}(aq;q)_{n+r}}.
 \label{beta2}
 \end{equation}
In addition
\begin{eqnarray}
 {\alpha^\prime}_r & = & \alpha_r\frac{(\rho_1;q)_r(\rho_2;q)_r(aq/\rho_1\rho_2)^r}
 {(aq/\rho_1;q)_r(aq/\rho_2; q)_r}, 
 \\
 \beta^\prime_n &=& \sum_{r\geq 0} \beta_r \frac{(\rho_1;q)_r(\rho_2;q)_r(aq/\rho_1\rho_2;q)_{n-r}(aq/\rho_1\rho_2)^j}
 {(q;q)_{n-j}(aq/\rho_1;q)_n(aq/\rho_2;q)_n}. 
\end{eqnarray}

{\bf Restricted products.}
We apply Bailey's transform in the case of restricted products.
\begin{eqnarray}
 u_{n-r}\! &:= &\! \prod_{m=0}^{n-r}(1-q^{m+1})\equiv (q;q)_{n-r} = \frac{(q;q)_\infty}{(qq^{n-r};q)_\infty}
 = \prod_{m=0}^\infty\frac{(1-q^{m+1})}{(1-q^{m+n-r+1})}
\\
& = & \frac{\cR(s=1-i\varrho(\tau))}{\cR (s=(n-r+1)(1-i\varrho(\tau)))},
\label{R1}
\\
v_{n+r}\! &:= &\! \prod_{m=0}^{n+r}(1-aq^{m+1})\equiv (aq;q)_{n+r} = \frac{(q;q)_\infty}{(aq^{n+r};q)_\infty}
= \prod_{m=0}^\infty\frac{(1-q^{m+1})}{(1-aq^{m+n+r+1})}
\\
& = & \frac{\cR(s=1-i\varrho(\tau))}{\cR(s=(n+r+\xi+1)(1-i\varrho(\tau)))},
\label{R2}
\end{eqnarray}
where $\xi = {\rm log}\,a/2\pi i\tau$.

Using Eqs. (\ref{R1}) and (\ref{R2}) we have:
\begin{eqnarray}
 \beta_n &=& \sum_{r=0}^n{\alpha_r}\frac{\cR(s=(n-r+1)(1-\varrho(\tau)))\cdot\cR(s=(n-r+1)(1-\varrho(\tau)))}
 {\cR(s=1-i\varrho(\tau))\cdot\cR(s=1-i\varrho(\tau))},
\\
\beta^{\prime}_n &=& \sum_{r=0}^n{\alpha^{\prime}_r}\frac{\cR(s=(n-r+1)(1-\varrho(\tau)))\cdot\cR(s=(n-r+1)(1-\varrho(\tau)))}
 {\cR(s=1-i\varrho(\tau))\cdot\cR(s=1-i\varrho(\tau))}.
\end{eqnarray}

In addition (see \cite{Andrews}),
\begin{eqnarray}
 {\alpha^\prime}_r  &=& \alpha_r \frac{(\rho_1;q)_r(\rho_2;q)_r(aq/\rho_1\rho_2)^r}
 {(aq/\rho_1;q)_r(aq/\rho_2;q)_r},
 \label{alpha}\\
 \beta^\prime_n &=& \sum_{r\geq 0}\beta_r \frac{(\rho_1;q)_r(\rho_2;q)_r(aq/\rho_1\rho_2;q)_{n-r}(aq/\rho_1\rho_2)^r}
 {(q;q)_{n-r}(aq/\rho_1;q)_n(aq/\rho_2;q)_n}. \label{beta}.
\end{eqnarray}

Important special cases of Bailey's Lemma and iterated of these results where found in \cite{Paule}.
Note that a pair of sequences $(\alpha_n, \beta_n)$ is called a {\it Bailey pair}. Thus if $(\alpha_n, \beta_n)$
is a Bailey pair, then   $(\alpha_{n}^\prime, \beta_{n}^\prime)$ is new pair which is given by
(\ref{alpha}) and (\ref{beta}).

{\bf Bailey's chain.}
Using Bailey initial pair $(\alpha^\prime_n, \beta^\prime_n)$ we can create new pair
$(\alpha^{\prime\prime}_n, \beta^{\prime\prime}_n)$ by applying Bailey's Lemma. Continuing this process we
create a sequence of Baley pairs -- Bailey's chain:
\begin{equation}
(\alpha_n, \beta_n) \longrightarrow (\alpha^\prime_n, \beta^\prime_n) \longrightarrow (\alpha^{\prime\prime}_n, \beta^{\prime\prime}_n)
\longrightarrow (\alpha^{\prime\prime\prime}_n, \beta^{\prime\prime\prime}_n)\longrightarrow \cdots
\label{chain1}
\end{equation}
Let us assume $(\alpha_n, \beta_n) = (\alpha_n^0, \beta_n^0)$, then this allows to extend the Bailey chain (\ref{chain1})
to the left \cite{Andrews}:
\begin{equation}
\cdots \longrightarrow (\alpha_n^{(-2)}, \beta_n^{(-2)}) \longrightarrow (\alpha^{(-1)}_n,
\beta^{(-1)}_n) \longrightarrow (\alpha^{0}_n, \beta^{0}_n)
\longrightarrow (\alpha^{\prime}_n, \beta^{\prime}_n)\longrightarrow \cdots
\label{chain}
\end{equation}

A Bailey pair is uniquely determined given either sequence $\alpha_n$ or $\beta_n$. Let $\alpha_n$ is given,
then $\beta_n$ sequence we can find. Eq. (\ref{beta1}) and Eq. (\ref{beta2}) can be invert, and result is \cite{Andrews84}:
\begin{eqnarray}
 \alpha_n &=& (1-aq^{2n})\sum_{k=0}^n \beta_k\frac{(aq;q)_{n+k-1}(-1)^{n-k}q^{C^{n-k}_2}}{(q;q)_{n-k}}
 \nonumber \\
&=& (1-aq^{2n})\sum_{k=0}^n \beta_k\frac{\prod_{m=0}^\infty (1-aq^{m+n+k})(-1)^{n-k}q^{C^{n-k}_2}}
{\prod_{m=0}^\infty(1-q^{m+n-k+1})}.
\end{eqnarray}
Finally we obtain
\begin{equation}
\alpha_n = (1-aq^{2n})\sum_{k=0}^n \beta_k\frac{\cR(s= (n+k)(1-i\varrho(\tau)))(-1)^{n-k}q^{C^{n-k}_2}}
{\cR(s= (n-k+1)(1-\varrho(\tau)))}\,.
\end{equation}

\subsection{Rogers-Ramanujan type identities}
\label{RR}

\begin{eqnarray}
\!\!\!\!\!&& \prod_{n=0}^\infty (1-2kq^n \cos \alpha + k^2q^{2n})  = (ke^{i\alpha}; q)_\infty(ke^{-i\alpha}; q)_\infty
 =\prod_{n=0}^\infty(1-ke^{i\alpha}q^n)\cdot(1-ke^{-i\alpha}q^n)
 \nonumber \\
\!\!\!\!\!&& =  \cR(s= \xi_+(1-i\varrho(\tau)))\cdot \cR(s= \xi_{-}(1-i\varrho(\tau))),
\\
\!\!\!\!\!&& \prod_{n=0}^\infty (1+2kq^n \cos \alpha + k^2q^{2n}) = (-ke^{i\alpha}q;q)_\infty(-ke^{-i\alpha}q;q)_\infty
= \prod_{n=0}^\infty(1+ke^{i\alpha}q^n)\cdot(1+ke^{-i\alpha}q^n)
\nonumber \\
\!\!\!\!\!&& = \cR(s= \xi_+(1-i\varrho(\tau))+i\sigma(\tau))\cdot \cR(s= \xi_{-}(1-i\varrho(\tau))+i\sigma(\tau)).
\end{eqnarray}
where $\xi_{\pm} = ({\rm log}\,k \pm i\alpha)/2\pi i\tau$.

\begin{eqnarray}
\theta_1(x) &=& 2\sum_{n=0}^\infty (-1)^n q^{(n+1/2)^2}\sin(2n+1)x
\nonumber \\
&=& 2q^{1/4}\sin x \prod_{n=1}^\infty(1-q^{2n})(1-2q^{2n}\cos 2x +  q^{4n}),
\end{eqnarray}

The Jacobi theta function $\theta_1(\tau, z)$ is
\begin{equation}
\theta_1(\tau, z) = i\sum_{n\in {\mathbb Z}}(-1)^n q^{\frac{1}{2}(n-\frac{1}{2})^2} y^{n-\frac{1}{2}}
= e^{2\pi i(1+\tau/8)}y^{-\frac{1}{2}}\prod_{n=1}^\infty(1-q^n)(1-yq^{n-1})(1-y^{-1}q^n),
\label{theta}
\end{equation}
where $q= e^{2\pi i\tau}$, $y= e^{2\pi i z}$.

\subsection{The {\it N=2} Landau-Ginzburg model}
\label{LG}

Elliptic genus in {\it N=2} superconformal field theory has been proposed by E. Witten \cite{Witten}
in order to understand the relation between {\it N=2} minimal models and Landau-Ginzburg theories.
In this section we discuss the basic properties satisfied by elliptic genera in {\it N=2} theories.
In the case of the {\it N=2} Landau-Ginzburg model an action takes the form \cite{Kawai}
\begin{equation}
\int d^2zd^2\theta d^2\overline{\theta}X\overline{X} + \left(\int d^2zd^2\theta W(X) + c.c.\,\, {term}\right),
\end{equation}
where the superpotential $W$ is a weighted homogeneous polynomial of $N$ chiral superfields
$X_1, \ldots, X_N$ with weights $\omega_1, \ldots, \omega_N$,
\begin{equation}
 \lambda W(X_1, \ldots, X_N) = W(\lambda^{\omega_1}X_1, \ldots, \lambda^{\omega_N}X_N)
\end{equation}

Assume that $W$ has an isolated critical point at the origin and $\omega_i\prime s$ are strictly positive
rational numbers such that $\omega_1,\ldots, \omega_N\leq 1/2$. then the elliptic genus of the
landau-Ginzburg model can be computed as
\begin{equation}
 Z(\tau, z) = \prod_{j=1}^N Z_{\omega_j}(\tau, z)
\end{equation}
where
\begin{equation}
Z_\omega(\tau, z)  =  \frac{\theta_1(\tau, (1-\omega)z)}{\theta_1(\tau, \omega z)}.
\end{equation}

Finally using Eqs. (\ref{theta})  we obtain:

\begin{eqnarray}
\!\!\!\!\!\!\!\!&& Z_\omega(\tau, z) = e^{\omega+\tau/8+1/2}
\prod_{n=1}^\infty\frac{[(1-yq^{n-1})(1-y^{-1}q^n)]\vert_{y=e^{2\pi i(1-\omega)z}}}
{[(1-yq^{n-1})(1-y^{-1}q^n)]\vert_{y=e^{2\pi i \omega z}}} = e^{\omega+\tau/8+1/2}
\nonumber \\
\!\!\!\!\!\!\!\! &&\times \frac{\cR(s= (1+z/\tau(1-\omega))(1-i\varrho(\tau)))
\cR(s=(1-z/\tau)(1-\omega)(1-\varrho(\tau)))}{\cR(s= (1+z\omega/\tau)(1-\varrho(\tau)))\cR(s=1-z\omega/\tau)(1-i\varrho(\tau))}.
\end{eqnarray}

{\bf The modular transformation laws.}

Obviously $\prod_{n=1}^\infty(1+q^n) = \prod_{n=1}^\infty (1-q^{2n})/(1-q^n) = \prod_{n=1}^\infty(1-q^{2n-1})^{-1}$,
thus
$
\cR(s=1-i\varrho(\tau)+i\sigma(\tau))\cR(s=1-\varrho(\tau)) = \cR(s= 1-2i\varrho(\tau)) = 1.
$

Next let us introduce some well-known functions and their modular properties under the action of
$SL(2, {\mathbb Z})$. The special cases associated with (\ref{R1}), (\ref{R2}) are:
\begin{eqnarray}
\varphi_1(q) & = & q^{-\frac{1}{48}}\prod_{m=1}^\infty
(1-q^{m+\frac{1}{2}})\,\, = q^{-\frac{1}{48}}\cR(s= 3/2(1-i\varrho(\tau)))\,\, \nonumber \\
& = & \frac{\eta_D(q^{\frac{1}{2}})}{\eta_D(q)}\,,
\\
\varphi_2(q) & = & q^{-\frac{1}{48}}\prod_{m=1}^\infty
(1+q^{m+\frac{1}{2}})\,\, = q^{-\frac{1}{48}}\cR(s=3/2(1-i\varrho(\tau))+i\sigma(\tau))\,\, \nonumber \\
& = &\frac{\eta_D(q)^2}{\eta_D(q^{\frac{1}{2}})\eta_D(q^2)}\,,
\\
\varphi_3(q) & = & \,\, \,\, q^{\frac{1}{24}}\prod_{m=1}^\infty
(1+q^{m+1})\,\, = q^{\frac{1}{24}}\cR(s= 2(1-i\varrho(\tau))+i\sigma(\tau))\,\,\nonumber \\
& = & \frac{\eta_D(q^2)}{\eta_D(q)}\,,
\end{eqnarray}
where
$
\eta_D(q) \equiv q^{1/24}\prod_{n=1}^\infty(1-q^{n})
$
is the Dedekind $\eta$-function. The linear span of $\varphi_1(q), \varphi_2(q)$
and $\varphi_3(q)$ is $SL(2, {\mathbb Z})$-invariant \cite{Kac_book}.

Functional equations for the spectral Ruelle functions are
\begin{eqnarray}
&&\!\!\!\!\!\!\!\!\!\!\!\!\!\!\!\!\!\!\!\! \cR(s= (z+b)(1-i\varrho(\tau))+i\sigma(\tau))\cdot\cR(s=-(1+z+b)(1-i\varrho(\tau))+i\sigma(\tau))
\nonumber \\
&&\!\!\!\!\!\!\!\!\!\!\!\!\!\!\!\!\!\!\!\! = q^{-zb-b(b+1)/2}\cR(s=-z(1-i\varrho(\tau))+i\sigma(\tau))\cdot
\cR(s=(1+z)(1-i\varrho(\tau))+i\sigma(\tau))
\nonumber \\
&&\!\!\!\!\!\!\!\!\!\!\!\!\!\!\!\!\!\!\!\! = q^{-z(b-1)-b(b+1)/2}\cR(s=(1-z)(1-i\varrho(\tau)+i\sigma(\tau))\cdot
\cR(s=z(1-i\varrho(\tau)+i\sigma(\tau))).
\label{Functional}
\end{eqnarray}
The simple case $b=0$ in Eq. (\ref{Functional}) leads to the symmetry $\tau \rightarrow -\tau$, i.e. the symmetry
$q\rightarrow q^{-1}$. The modular transformation laws are:

$\theta_1(\tau+1, z)= e^{2\pi i(1/8)}\theta_1(\tau,z)$; $\theta_1(-1/\tau, z/\tau) =
(-i\tau)^{1/2}e^{2\pi i(1/2)(z^2/\tau)}\theta_1(\tau, z)$, $\theta_1(\tau,-z) = -\theta_1(\tau, z))$.

The double quasi-periodicity law is: $\theta_1(\tau, z+\alpha\tau +\beta) =
(-1)^{\alpha+\beta}e^{-2\pi i(1/2)(\alpha^2 +2\alpha z)}\theta_1(\tau,z)$, \,\,$\alpha, \beta \in {\mathbb Z}$.
Note that $\theta_1(\tau, z)$ has no poles but this function (as a function of $z$) has simple zeros:
$\theta_1(\tau, \alpha\tau +\beta)=0,\,\, \alpha, \beta \in {\mathbb Z}$.

\section*{Asknowledgments}

AAB and AEG would like to acknowledge the Conselho Nacional
de Desenvolvimento Cient\'{i}fico e Tecnol\'{o}gico (CNPq, Brazil) and
Coordenac\~{a}o de Aperfei\c{c}amento de Pessoal de N\'{i}vel Superior
(CAPES, Brazil) for financial support.


\begin{thebibliography}{999999999}



\bibitem{Schur}
I. Schur, {\it \"{U}ber eine Klasse von Matrizen die sich einer gegebenen Matrix zuordnen
lassen}. Ges. Abhandlung {\bf 1} (1901) 1.

\bibitem{Williams}
P. Perry and F. Williams, {\it Selberg zeta function and trace formula for the BTZ black hole},
Int. J. Pure Appl. Math. {\bf 9} (2003) 1-21.

\bibitem{Bytsenko}
A. A. Bytsenko, M. E. X. Guim\~{a}raes and F. L. Williams, {\it Remarks on the spectrum and
truncated heat kernel of the BTZ black hole}, Lett. Math. Phys. {\bf 79} (2007) 203-211.

\bibitem{BB}
L. Bonora and A.A. Bytsenko, {\it Partition functions for quantum gravity, black holes, elliptic
genera and Lie algebra homologies}, Nucl. Phys. B {\bf 852} (2011) 508-537.

\bibitem{Ray}
D. Ray and I. Singer, {\it R-Torsion and the Laplacian on Riemannian Manifolds}, Adv. Math.
{\bf 7} (1971) 145-210.

\bibitem{Farber}
M. Farber and V. Turaev, {\it Poincar\'{e}-Reidemaister Metric, Euler Structures, and Torsion},
J. reine angew. Math. {\bf 520} (2000) 195-225.

\bibitem{Fried}
D. Fried, {\it Analytic torsion and closed geodesics on hyperbolic manifolds}, Invent.
Math. {\bf 84} (1986) 523-540.

\bibitem{Bytsenko97}
A. A. Bytsenko, L. Vanzo and S. Zerbini, {\it Ray-Singer torsion for a hyperbolic 3-manifold and
asymptotics of Chern-Simons-Witten invariant}, Nucl. Phys. B {\bf 505} (1997) 641-659.

\bibitem{Witten}
E. Witten {\it On the Landau-Ginzburg description of N=2 minimal models}, Int. J. Mod. Phys. A
{\bf 9} (1994) 4783-4800.
\bibitem{Macdonald}
I. G. Macdonald, {\it Symmetric Functions and Hall Polynomials}, 2nd Ed. Clarendon Press,
Oxford, 1995.

\bibitem{fauser:jarvis:2003a}
B. Fauser and P. D. Jarvis,
{\it A Hopf laboratory for symmetric functions}, J. Phys. A: Math. Gen. {\bf 37} ( 2004)
1633-1663.


\bibitem{fauser:jarvis:king:wybourne:2006a}
B. Fauser, P. D. Jarvis, R. C. King and B. G. Wybourne, {\it New branching rules induced
by plethysm}, J. Phys A: Math. Gen. {\bf 39} (2006) 2611-2655.


\bibitem{Fauser10}
B. Fauser, P. D. Jarvis and R. C. King, {\it Plethysms, replicated Schur functions and series,
with applications to vertex operators}, J. Phys. A: Math. Theor. {\bf 43} (2010) 405202 (30 pp).

\bibitem{littlewood:richardson:1934a}
D. E. Littlewood and A. Richardson,
{\it Group characters and algebra}, Phil. Trans. Roy. Soc. London A {\bf 233} (1934) 99-141.

\bibitem{littlewood:1950a}
D. E. Littlewood,
{\it The Theory of Group Characters}, 2nd Ed. Oxford University Press, Oxford, 1950.

\bibitem{buch:2000a}
A. S. Buch, {\it The saturation conjecture (after A. Knutson and T. Tao). With an
appendix by W. Fulton}, Enseign. Math. {\bf 46} (2000) 43-60.


\bibitem{littlewood:1950b}
D. E. Littlewood, {\it A University Algebra}, Review, Bull. Amer. Math. Soc. {\bf 59}
(1953) 97-98.

\bibitem{chen:garsia:remmel:1984a}
Y. Chen, A. M. Garsia and J. Remmel,
{\it Algorithms for plethysm}, Contemporary Math. {\bf 34} (1984) 109-153.

\bibitem{Andrews}
G. E. Andrews, {\it q-Series: Their Development and Application in Analysis, Number Theory, Combinatorics,
Physics, and Computer Algebra}, Expository Lectures from the CBMS Regional Conference No. 66, Providence,
Rh. I.: AMS, 1986.

\bibitem{Drissi}
L. B. Drissi, J. Houda and E. H. Saidi, {\it Refining the Shifted Topological Vertex}, J. Math. Phys.
{\bf 50} (2009) 0135 ; arXiv:0812.0513 [hep-th].

\bibitem{GopakumarI}
 R. Gopakumar and  C. Vafa, {\it M-theory and topological strings-I},
 hep-th/9809187.

\bibitem{GopakumarII}
R. Gopakumar and C. Vafa, {\it M-theory and topological strings-II},
hep-th/9812127.

\bibitem{Gopakumar99}
R. Gopakumar and C. Vafa, {\it On the gauge theory/geometry correspondence},
Adv. Theor. Math. Phys. {\bf 3} (1999) 4115-1443.


\bibitem{Hollowood}
T. J. Hollowood, A. Iqbal and C. Vafa, {\it Matrix Models,
Geometric Engineering and Elliptic Genera}, JHEP 0803 (2008) 069;
[arXiv:hep-th/0310272v4].

\bibitem{Antoniadis}
I. Antoniadis, E. Gava, K. S. Narain and T. R. Taylor, {\it Topological amplitudes in
string theory}, Nucl. Phys. B {\bf} 413 (1994) 162

\bibitem{Bershadsky}
M. Bershadsky, S. Cecotti, H. Ooguri and C. Vafa, {\it Kodaira-Spencer theory of gravity
and exact results for quantum string amplitudes}, Commun. Math. Phys. {\bf 165} (1994) 311.


\bibitem{Bailey}
W. N. Bailey {\it Identities of the Rogers-Ramanujan type}, Proc. London Math. Soc.
{\bf 50} (1949) 1-10.

\bibitem{Paule}
P. Paule {\it On identities of the Rogers-Ramanujan type}, J. Math. Anal. and Appl.
{\bf 107} (1985) 255-284.

\bibitem{Andrews84}
G. E. Andrews, {\it Multiple series Rogers-Ramanujan type identities},
Pacific J. Math. {\bf 114} (1984) 267-283.

\bibitem{Kawai}
T. Kawai, Y. Yamada and S.-K. Yang, {\it Elliptic Genera and N=2 Superconformal Field Theory},
Nucl. Phys. B {\bf 414} (1994) 191-212.

\bibitem{Kac_book}
V. G. Kac, {\it Infinite dimensional Lie algebras}, Cambridge University Press, 1990.





\end{thebibliography}
\end{document}